\documentclass[conference]{IEEEtran}
\IEEEoverridecommandlockouts
\usepackage{cite}
\usepackage{multicol}
\usepackage{balance}
\usepackage{changes}
\usepackage{amsmath,amssymb,amsfonts}
\usepackage{algorithmic}
\usepackage{tablefootnote}
\usepackage{graphicx}
\usepackage[nolist]{acronym}
\usepackage{multirow}
\usepackage{array}
\usepackage{textcomp}
\usepackage[hidelinks]{hyperref}
\usepackage{url}
\usepackage{changes}
\usepackage{float}
\usepackage[numbers, sort&compress]{natbib}
\usepackage{pifont}
\usepackage[margin=2cm]{geometry} 
\usepackage{caption}

\usepackage{xcolor}
\def\BibTeX{{\rm B\kern-.05em{\sc i\kern-.025em b}\kern-.08em
    T\kern-.1667em\lower.7ex\hbox{E}\kern-.125emX}}

\usepackage{enumitem}
\setlist[itemize]{leftmargin=10pt}
   
\begin{document}

\title{Adaptive Cyber-Attack Detection in IIoT Using Attention-Based LSTM-CNN Models}

\makeatletter 
\newcommand{\linebreakand}{%
  \end{@IEEEauthorhalign}
  \hfill\mbox{}\par
  \mbox{}\hfill\begin{@IEEEauthorhalign}
}
\makeatother 

\author{\IEEEauthorblockN{Afrah Gueriani}
\IEEEauthorblockA{\textit{LSEA Lab., Faculty of Technology} \\
\textit{University of MEDEA}\\
Medea 26000, Algeria\\
gueriani.afrah@univ-medea.dz }
\and
\IEEEauthorblockN{Hamza Kheddar}
\IEEEauthorblockA{\textit{LSEA Lab., Faculty of Technology} \\
\textit{ University of MEDEA}\\
Medea 26000, Algeria \\
kheddar.hamza@univ-medea.dz}
\and 
\IEEEauthorblockN{Ahmed Cherif Mazari}
\IEEEauthorblockA{\textit{LSEA Lab, Faculty of Science} \\
\textit{ University of MEDEA}\\
Medea 26000, Algeria \\
mazari.ahmedcherif@univ-medea.dz}
}

\makeatletter

\def\ps@headings{%
\def\@oddhead{\parbox[t][\height][t]{\textwidth}{\flushleft

\noindent\makebox[\linewidth]
}
\vspace{0.5cm}
\hfil\hbox{}}%
\def\@oddfoot{\MYfooter}%
\def\@evenfoot{\MYfooter}}
\def\ps@IEEEtitlepagestyle{%
\def\@oddhead{\parbox[t][\height][t]{\textwidth}{
2024 International Conference on Telecommunications and Intelligent Systems (ICTIS)\\

}\hfil\hbox{}}%

\def\@oddfoot{ 979-8-3315-2739-6/24/\$31.00 \textcopyright 2024 IEEE \hfil 
\leftmark\mbox{}}%
\def\@evenfoot{\MYfooter}}

\maketitle

\begin{abstract}
The rapid expansion of the industrial Internet
of things (IIoT) has introduced new challenges in securing critical infrastructures against sophisticated cyberthreats. This study presents the development and evaluation of an advanced Intrusion detection (IDS) based on a hybrid LSTM-convolution neural network (CNN)-Attention architecture, specifically designed to detect and classify cyberattacks in IIoT environments. The research
focuses on two key classification tasks: binary and multi-class classification. The proposed models was rigorously tested using the Edge-IIoTset dataset. To mitigate the class imbalance in the dataset, the synthetic minority over-sampling technique (SMOTE) was employed to generate synthetic samples for the underrepresented classes. This ensured that the model could learn effectively from all classes, thereby improving the overall classification performance. Through systematic experimentation, various deep learning (DL) models were compared, ultimately demonstrating that the LSTM-CNN-Attention model consistently outperformed others across key performance metrics. In binary classification, the model achieved near-perfect accuracy, while in multi-class classification, it maintained a high accuracy level (99.04\%), effectively categorizing different attack types with a loss value of 0.0220\%.

\end{abstract}


\begin{acronym}
 \acro{IDS}{intrusion detection system} \acro{CNN}{convolution neural network} 
\end{acronym}

\section{Introduction}
\label{sec1}

The Internet of Things (IoT) has emerged as a transformative technology, facilitating a vast network of interconnected devices that collect, exchange, and analyze data in real-time \cite{mishra2021internet,aldhaheri2023deep, kumari2023comprehensive,aguru2024lightweight}. These devices communicate seamlessly, with end devices, often referred to as sensor nodes, continuously monitoring their environment and transmitting alert messages to other connected devices \cite{mondal2020secure}. This technological revolution is reshaping industries, from smart homes and healthcare to agriculture and manufacturing, by driving efficiency, innovation, and automation \cite{gueriani2024enhancing,jaidka2020evolution}. Within this broader context, Industrial Internet
of Things (IIoT) represents a specialized subset of IoT \cite{sithungu2022gaainet}, focused on integrating advanced connectivity, machine learning (ML), and big data analytics into industrial processes. The IIoT enhances operational performance and enables predictive maintenance, energy optimization, and real-time monitoring, thereby creating smarter and more responsive industrial systems. However, the widespread adoption of IIoT has also introduced significant cybersecurity challenges \cite{sithungu2022gaainet}. The extensive inter-connectivity and the critical nature of industrial operations make IIoT environments attractive targets for cyber-attacks. Traditional cybersecurity measures, designed for more static and less interconnected systems, are often inadequate for the dynamic and complex nature of IIoT networks. These networks generate large volumes of heterogeneous data, exhibit temporal dependencies, and are often distributed across geographically dispersed locations. Such characteristics pose unique challenges for \ac{IDS} tasked with safeguarding these environments. This later is one of the solutions to limit these suspicious is \ac{IDS}, which aims to identify potential security threats to a network \cite{wang2023securing,10729241}. Artificial intelligence (AI) models especially DL methods have been used to properly solve malware infections and cybersecurity problems like in \cite{do2023using}. 

Among DL algorithms, the long short-term memory (LSTM), \ac{CNN} \cite{altunay2023hybrid}, and Attention mechanism \cite{ullah2023abdnn} are employed in this paper. LSTM, a special class of recurrent neural network (RNN).  CNN, on the other hand, excel at identifying spatial hierarchies and local patterns, which are crucial for detecting anomalous behaviors in network traffic and sensor data. The attention layer, a key component in transformers, enhances these capabilities by enabling sequence-to-sequence mappings \cite{kheddar2024transformers,habchi2024machine,djeffal2023automatic}. 
Attention-based LSTM-\acp{CNN} networks combine the temporal modeling strengths of LSTMs, the spatial feature extraction capabilities of \acp{CNN}, and the dynamic focus provided by attention mechanisms. 

This research contributes to the development of robust and resilient security solutions for IIoT by presenting an attention mechanism-based LSTM-\ac{CNN} model that ensures the protection and integrity of industrial systems in an increasingly connected world. The following steps outline the procedural flow and the contributions of this paper:

\begin{itemize}
\item Propose a range of distinct models for detecting malware in IIoTs. Specifically, the proposed models include LSTM, \ac{CNN}, and hybrid LSTM-\ac{CNN} architectures, with and without attention layer.

\item Evaluate the proposed models by utilizing the Edge-IIoTset dataset and applying various metrics to determine the effectiveness of the architecture.
 \item Select the most effective model (\#10 LSTM-\ac{CNN}-Attention model) for multiclass classification by evaluating its performance based on accuracy and loss. Additionally, provide a detailed analysis using other relevant metrics, for both binary and multi-class classification tasks.

\item Compare the proposed methodology to existing state-of-the-art methods using the same datasets.
\end{itemize}

The remainder of the paper is organized as follows: Section \ref{sec2} presents the preliminaries, including a review of related literature. Section \ref{sec3} describes the proposed methodology and the dataset preprocessing steps. In Section \ref{sec4}, we discuss the experimentation, results, and analysis. Finally, Section \ref{sec5} concludes the paper and outlines future research directions.

\section{Literature review}
\label{sec2}

DL models have emerged as powerful tools with demonstrated effectiveness across diverse domains, including cybersecurity. Their superior performance over traditional machine learning (ML) techniques underscores their utility in tackling complex challenges in this field. \cite{alshehri2024self} leverages self-attention mechanisms integrated with deep convolutional neural networks (SA-\acp{CNN}) to enhance the detection accuracy of cyber-attacks. The self-attention layer helps the model focus on the most relevant features, improving its ability to identify malicious activities within complex IIoT environments. The performance of the SA-DCNN model is evaluated using both the IoTID20 and Edge-IIoTset datasets, with results showing significant improvements in detection performance, showcasing the potential of self-attention in IIoT security.

Moving on,  \cite{de2022hybrid} presents a hybrid approach using CNN and LSTM models for intrusion detection in IIoT environments. This model is designed to operate at the edge, ensuring data privacy while detecting attacks. By combining CNN for feature extraction and LSTM for temporal pattern recognition, the model effectively identifies complex intrusions using features exclusively from the transport and network layers. The approach improves detection accuracy while maintaining data confidentiality, demonstrating promising results in edge-based IIoT security applications.

Gueriani et al. \cite{gueriani2024enhancing} proposes an \ac{IDS} specifically tailored for IoT environments, using a hybrid approach combining CNN and LSTM DL methods to accurately distinguish between normal and anomalous IoT traffic. The authors validate their approach using two benchmark datasets: the CICIoT2023 dataset (comprising two subsets) and the CICIDS2017 dataset. Results demonstrate that this approach significantly outperforms current state-of-the-art techniques in ML and DL in terms of accuracy and quality. Similarly, \cite{sharma2024explainable} proposes an \ac{IDS} specifically designed to enhance the security of IoT networks, leveraging a hybrid approach combining CNN and LSTM DL methods to accurately distinguish between normal and anomalous IoT traffic. The authors validate their approach using the CICIoT2023 dataset (comprising two subsets) and the CICIDS2017 dataset. The results demonstrate that this approach significantly outperforms current state-of-the-art ML and DL techniques in terms of accuracy and quality. The study \cite{ullah2023magru} introduces the MAGRU model, which combines multi-head attention mechanisms with gated recurrent units (GRUs) to effectively detect and classify cyber-attacks. The proposed model is evaluated on the Edge-IIoTset and MQTTset datasets, demonstrating superior performance in terms of accuracy and detection rate compared to traditional models, making it a promising solution for enhancing security in IIoT environments.

\section{proposed methodology}
The purpose of this section is to discuss the attention mechanism and describe our suggested model, an LSTM-CNN-Attention DL architecture. This approach is intended to identify and categorize hostile and benign activity in a real-environmental dataset. The following stages are included in the suggested plan:

\noindent \textbf{\textit{- Attention mechanism:}} An attention function can be understood as a mechanism that maps a query and a set of key-value pairs to an output, with the query, keys, values, and output all represented as vectors. The output is computed as a weighted sum of the values, where the weights are derived from a compatibility function that assesses the relationship between the query and each corresponding key \cite{vaswani2017attention}. For a detailed description of the attention mechanism, refer to Equation \ref{eq11} in \cite{kheddar2024automatic}:

         \begin{equation}
     \small
        \mathrm{Attention (Q, K, V)}= softmax\left(\frac{Q \times K^T}{\sqrt{d_{q}}}\right) \times V
        \label{eq11}
    \end{equation}

Where,  \textit{Q (Query)} is the input that the model uses to compute attention scores. \textit{d\_{q}} is the length of Q. \textit{K (Key)} is the set of possible matches against which the query is compared. \textit{V (Values)} represent the information the model utilizes after calculating attention scores, based on the corresponding keys.

In this paper, the attention mechanism improves cyber-attack detection by emphasizing relevant features, enhancing temporal understanding, and isolating anomalies, outperforming conventional CNN-LSTM architectures in IIoT environments.


\label{sec3}
\noindent \textbf{\textit{- Data preprocessing:}} In the initial phase, the dataset, containing 60 features and 15 attack types, was structured into a two-dimensional format, with labels converted to binary for model compatibility.

\noindent \textbf{\textit{- Data splitting:}} The dataset was split 80\% for training and 20\% for testing, using class weights to address imbalance. The Adam optimization algorithm was employed for efficient convergence and optimal performance.

\noindent \textbf{\textit{- Data numerization:}} A numerization approach was used to make processing of non-numeric categorical data easier. In particular, non-numeric categories were transformed into numeric values using the "LabelEncoder" approach in combination with the "fit\_transform" method, making the data suitable for further analysis and modeling.

\noindent \textbf{\textit{- Addressing class imbalance with SMOTE:}} The SMOTE algorithm is employed to mitigate class imbalance by creating synthetic samples for the minority class. Given that SMOTE requires a 2D input format, the feature matrix is initially flattened to a 2D structure with dimensions \textbf{(samples, features)} before applying the technique. Once the synthetic data is generated, the matrix is reshaped back to its original 3D form to maintain compatibility with the model. After the application of SMOTE, the dataset is divided into training and validation sets once again. The balanced distribution of classes is then confirmed by printing the resampled dataset shape using the command \textbf{print(f"Resampled dataset shape: Counter(y\_resampled)")}, which verifies that the SMOTE algorithm has successfully equalized the class distribution. \\
\noindent \textbf{\textit{- Model architecture:}} Multiple CNN, LSTM, and LSTM-CNN models with and without an attention layer were utilized in the formulation of our suggested model. We used a variety of layers, including an input layer, LSTM layers, batchnorm., dropout, attention layer, and danse layer. \ref{fig2}. 

\begin{figure}[h!]
\centering
\includegraphics[scale=0.6]{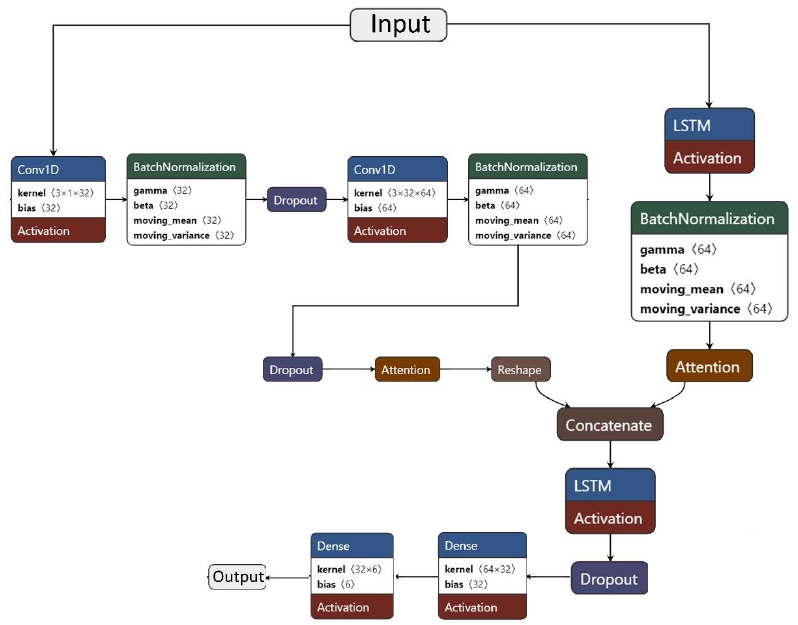}
\caption{The proposed LSTM-CNN-Attention model.}
\label{fig2}
\end{figure} 

First, the model receive an \textbf{Input layer} which represents the input data, The first bloc is the CNN bloc, we have two \textbf{Conv1D layer} use filters to extract local features from the input data. These layers learn patterns in the data through "kernel" weights.
"Bias" terms adjust the output of the filters.
The "activation" function introduces non-linearity, making the model capable of learning more complex relationships. a \textbf{Batch Normalization layer} is also used in our model, this layer normalizes the activations of the previous layer. This helps to prevent the vanishing gradient problem and improves the performance of the model. Then we have \textbf{Dropout layer} which helps to prevent overfitting and improves the generalization ability of the model. The \textbf{Attention mechanism} is then applied, it is a key aspect of our model. It allows the model to focus on the most important parts of the input sequence at each time step, and selectively attends to certain parts of the data, improving the model's ability to learn relevant information. a \textbf{Reshape layer} is used after that. The second branch of our model concerns the \textbf{LSTM layers}. The model uses two LSTM layers which have internal "memory" mechanisms that allow them to capture long-range dependencies in the data. "Activation" introduces non-linearity after each LSTM layer followed by a \textbf{BatchNormalization} and \textbf{Attention} layers. The outputs from the two branches are then merged through a \textbf{Concatenation layer}, allowing the model to simultaneously capture both long-term dependencies and the most prominent features of the input. The last layers were two  \textbf{Dense (fully connected) layers}, these layers perform further processing, connecting all neurons in one layer to all neurons in the next.
They learn more complex combinations of features, moving towards the final output. The size of this output layer is two for binary classification and six for multiclass classification.


\section{Experimentation, results and discussion}
\label{sec4}


\subsection{Dataset exploration}
The effectiveness of our proposed workflow was evaluated using the Edge-IIoTset dataset\footnote{\url{ https://www.kaggle.com/code/mohamedamineferrag/edge-iiotset-pre-processing}}, a realistic benchmark specifically designed to facilitate the development of security analytics applications for real-world IoT and industrial IoT operations. This dataset, widely employed in \acp{IDS} incorporating ML approaches, comprises a diverse range of attack scenarios, 
as illustrated in Figure 3. The dataset was generated using a heterogeneous array of IoT devices, including over ten different types, such as low-cost digital sensors for temperature and humidity, pH sensor meters, ultrasonic sensors, heart rate sensors, water-level detection sensors, soil moisture sensors, flame sensors, and others. Notably, the dataset features 61 novel attributes out of a total of 1176 characteristics, providing a comprehensive and nuanced representation of IoT traffic 
\cite{ferrag2022edge}.

\begin{figure}[h!]
\centering
\includegraphics[scale=0.13]{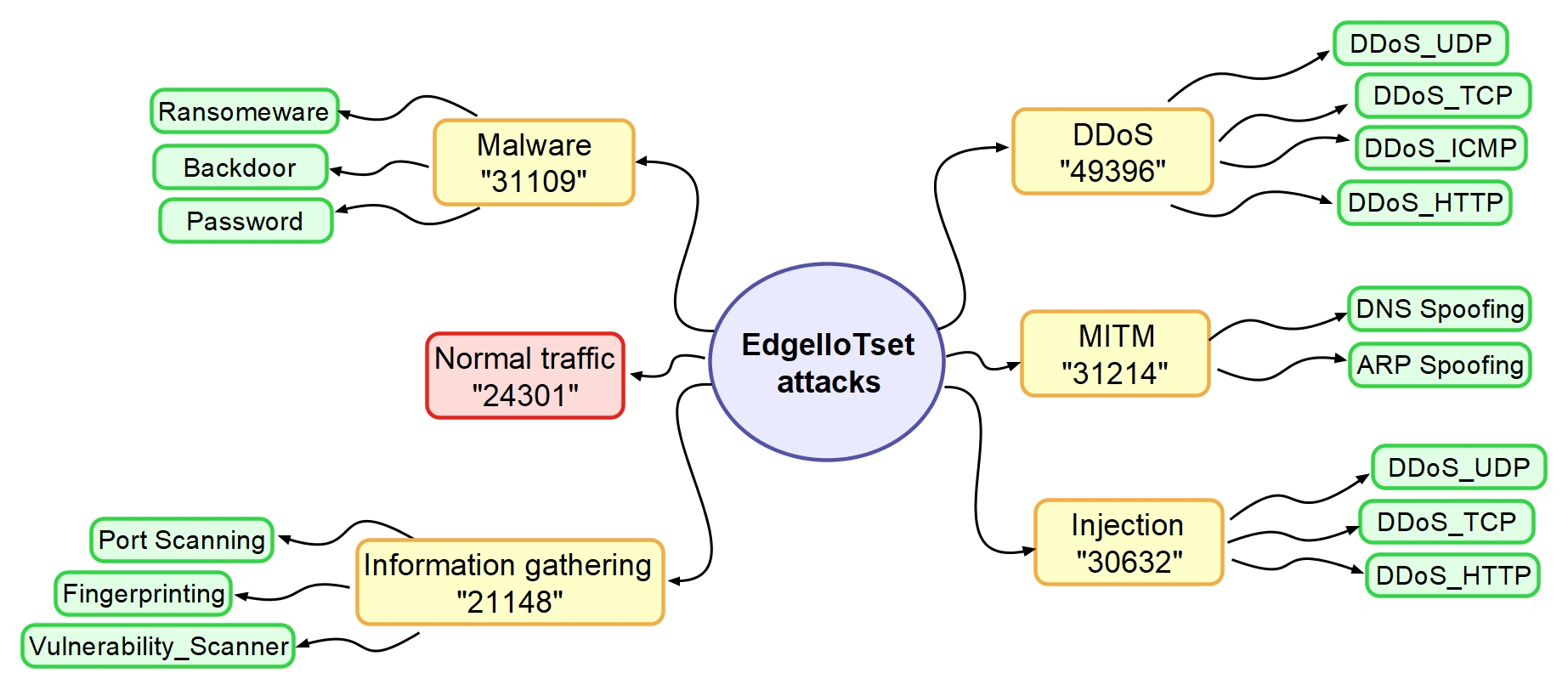}
\caption{The Edge-IIoTset dataset attacks types with the number of samples.}
\label{fig3}
\end{figure}

\subsection{Performance metrics}
To thoroughly evaluate the performance of our LSTM architecture, we utilized various metrics. 
These metrics, which are extensively documented and defined in sources such as \cite{gueriani2023deep, gueriani2024enhancing, kheddar2024deep, dunn2020robustness}, were derived from the four fundamental classifications: true positive (TP), true negative (TN), false positive (FP), and false negative (FN). Specifically, TP and TN represent the correctly classified instances of legitimate and attack vectors, respectively, while FP and FN denote the incorrectly classified legitimate and attack vectors. The following are the matching equations \cite{kheddar2023deep}:

    \begin{equation}
    \scriptsize
        \mathrm{Acc}=\mathrm{\frac{TP+TN}{TP+FP+TN+FN}}
        \label{eq1}
    \end{equation} 

    \begin{equation}
    \scriptsize
        \mathrm{Rc}=\mathrm{\frac{TP}{TP+FN}}, \hspace{0.5cm}  \mathrm{Pr}=\mathrm{\frac{TP}{TP+FP}}
        \label{eq2}
    \end{equation}

    \begin{equation}
    \scriptsize
        \mathrm{F1-Score}= \mathrm{2 \times \frac{Precision \times Recall}{Precision + Recall}}
        \label{eq6}
    \end{equation}

         \begin{equation}
     \scriptsize
        \mathrm{FPR}= \mathrm{\frac{FP}{FP+TN}}
        \label{eq5}
    \end{equation}

\subsection{Experiments and Results}
In this subsection, we present the experimental results evaluating the performance of the proposed models. The evaluation was conducted using the Edge-IIoTset dataset and various performance metrics. Model training was performed on Google Colab using a GPU, with the Adam optimizer over the course of 19 epochs.

\noindent \textbf{\textit{{- Model selection strategy:}}} 
Table \ref{models} shows the performance of different model variants for multiclass classification before and after applying SMOTE technique. The model variants are compared based on their accuracy and loss, taking into account the results after balancing the dataset using the SMOTE technique.
\begin{itemize}
    \item The models that use attention mechanisms, such as LSTM-Attention (\#3), generally have better accuracy and lower loss than the models that do not use attention (\#1 and \#2). 
\end{itemize}
\begin{itemize}
 \item LSTM without and with attention mechanism (\#2 and \#3 respectively) and LSTM-CNN (\#4 to \#9)  models perform much better than CNN (\#1). This is likely because LSTMs are better at capturing temporal dependencies in the data, which can be important for attack detection.
\end{itemize}

\begin{itemize}
 \item Adding attention mechanisms to LSTM-CNN (\#6 to \#9) models can further improve performance. Attention mechanisms allow the model to focus on the most important parts of the input, which can help to improve accuracy.
\end{itemize}

\begin{itemize}
 \item The best performing model is the LSTM-CNN-Attention model with a Dropout layer (\#9). Dropout is a regularization technique that helps to prevent overfitting, which can improve generalization performance.
\end{itemize}

Table \ref{models} shows that LSTM-CNN models with attention and dropout are very effective for multi-class classification tasks, including attack detection.

\noindent \textbf{\textit{{- Accuracy and loss graph:}}} 
The graphs in Figure \ref{fig4} shows the 
accuracy of binary and multiclass classification (a) and loss for binary and multiclass classification task (b). The binary accuracy of attack detection in (a) is very high, reaching 100\% after the 2nd epoch and remaining there throughout the training. This indicates that the model quickly learned to distinguish between legitimate and malicious data, achieving near-perfect accuracy in the process. Concerning the multiclass classification accuracy, the model demonstrates stable validation accuracy, consistently reaching around 99.04\%.
The binary loss in Figure \ref{fig4} (b) starts at a high value of 0.04 and drops drastically in the first two epochs before stabilizing at a very low value, around 1.99e-07. Additionally, the loss continues to decrease, with the validation loss stabilizing at 0.0220\%. These results indicate a well-trained and robust model, exhibiting strong performance in the multiclass classification of attacks.

\begin{figure}[h!]
\centering
\includegraphics[scale=0.5]{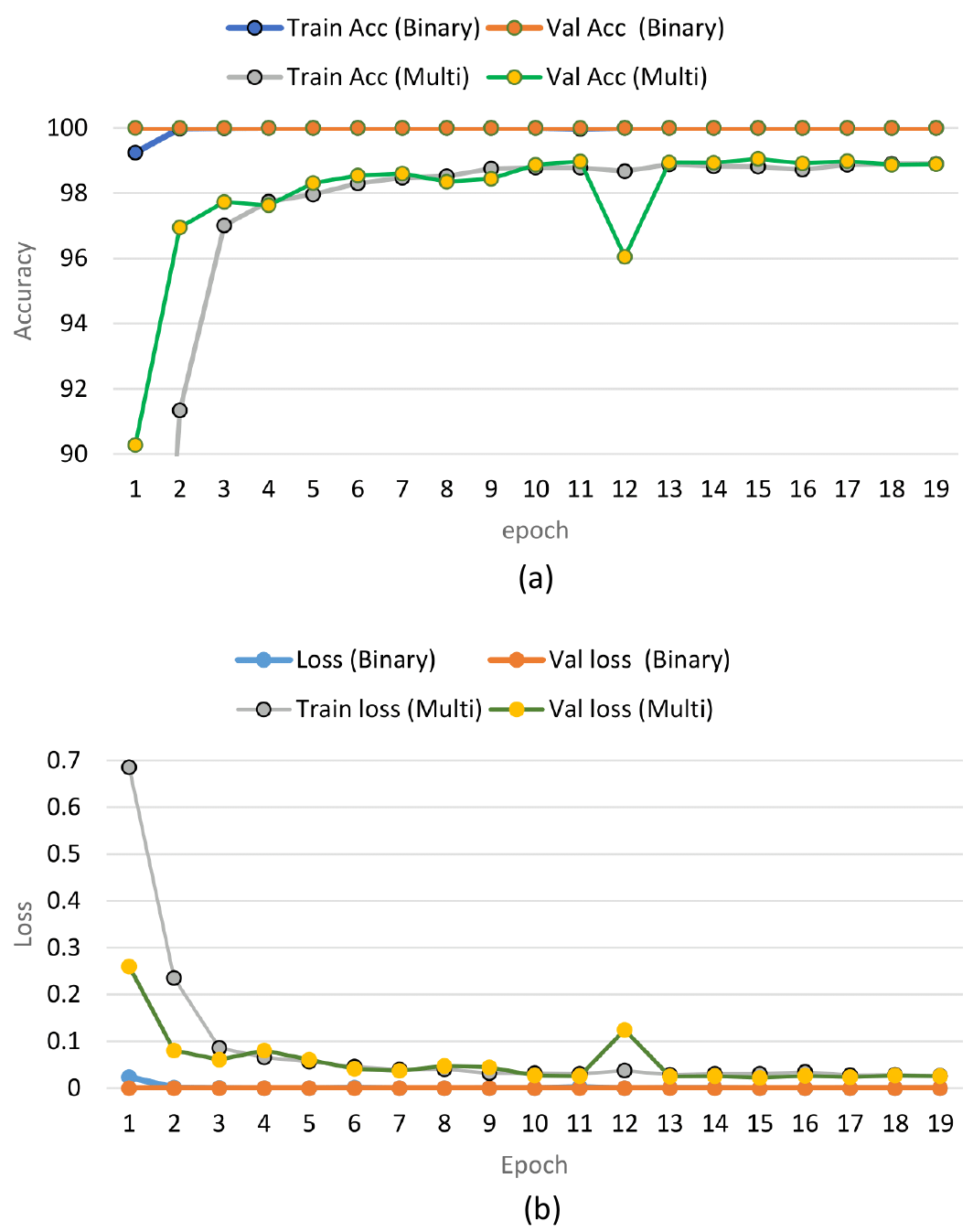}
\caption{Accuracy and loss of the LSTM-CNN-Attention: (a) Accuracy of binary and multi-class classification, (b) Loss of binary and multi-class  classification.}
\label{fig4}
\end{figure}


\begin{table}[h]
\caption{Detection accuracy and loss across various model variants in multiclass classification.}
\label{tab1}
\scriptsize
\centering
\begin{tabular}{@{}m{0.3cm}m{2.3cm}m{1cm}m{1cm}m{1cm}m{1cm}@{}}
\hline 
&  & \multicolumn{2}{c}{Before applying SMOTE} & \multicolumn{2}{c}{After applying SMOTE}   \\
\hline 
Index & Architecture variant & Acc. (\%) & Loss (\%) &  Acc. (\%)  & Loss (\%) \\
\hline
\#1 & CNN & 95.60 & 0.0335 & 94.99 & 0.1210\\[1mm]


\#2 & LSTM & 98.68 & 0.0394 & 98.68 & 0.0340\\[1mm]

\#3 & LSTM-Atten. & 98.13 & 0.0167 & 98.77 & 0.0326\\[1mm]

\#4 & LSTM-CNN with 2 Dense layers & 98.58 & 0.0144 & 98.64 & 0.0357\\[1mm]

\#5 & LSTM-CNN-Atten. with 2 Dense layers & 98.59 & 0.0138 & 98.58 & 0.0358\\[1mm]

\#6 & LSTM-CNN-Attention with 3 Dense layers & 98.57 & 0.0148 & 98.80 & 0.0339\\[1mm]

\#7 & LSTM-CNN-Atten. with MaxPooling & 97.84 & 0.0219 & 98.87 &0.0288\\[1mm]

\#8 & (LSTM-CNN)-Atten. & 98.76 & 0.0123 & 98.62 & 0.0355\\[1mm]

\#9 & (LSTM-CNN)-Atten. added a DropOut layer & 98.80 & 0.0114 & 99.04 & 0.0220\\[1mm]

\hline
\end{tabular}
\label{models}
\end{table}

\noindent \textbf{\textit{{- Classification report:}}} The binary classification results indicate perfect accuracy, precision, recall, and F1-score for both attack and normal traffic, with no misclassifications. Table \ref{Mc} presents the multiclass classification performance of the LSTM-CNN model with attention, showing high accuracy and F1-scores across all classes. The model achieves perfect scores for "Normal traffic," "Info. gathering," "MITM," and "Injection." Slightly lower performance is observed for "DDoS" and "Malware," with F1-scores of 97\%, suggesting minor challenges in classifying these classes. Overall, these high scores highlight the model’s potential for real-world network security applications.


\textbf{The FPR} of binary classification achieved an excellent value of 0\%.  This means that the model correctly identified all negative cases, indicating very high accuracy in detecting TN.  In mult-iclass classification, the FPR slightly increased to 0.002\%, still extremely low. This small rise, due to the complexity of mult-iclass tasks, demonstrates the model's robustness and reliability.

\begin{center}
\begin{table}
\centering
\caption{Multiclass classification report for LSTM-CNN-Attention model.}
\label{Mc}
\begin{tabular}{llllllll}
\hline
& Precision & Recall & F1-score & VD\\ [0.4mm]
\hline
Normal traffic (0) & 100\% & 100\% & 100\% &9860 \\[0.6mm]

DDoS (1) & 98\% & 96\% & 97\% & 10061\\ [0.4mm]

Info. gathering (2) & 100\% & 100\% & 100\% & 9869\\[0.4mm]

MITM (3) & 100\% & 100\% & 100\% & 9815 \\ [0.4mm]

Injection (4) & 100\% & 100\% & 100\% & 9944\\[0.4mm]

Malware (5) & 96\% & 98\% & 97\% & 9727\\ [0.4mm]
\hline

Accuracy &  &  & 99\% & \\[0.4mm]

Macro avg & 99\% & 99\% & 99\% & \\[0.4mm]

 Weighted avg & 99\% & 99\% & 99\% & \\[0.4mm]

  Total VD & & &  & 59276\\[0.4mm]
\hline
\end{tabular}
\begin{flushleft}
\vspace{0.08cm}
 Abbreviation: Validation data (VD)   
\end{flushleft}
\end{table}
\end{center}

\noindent \textbf{\textit{{- Confusion matrix:}}} The LSTM-CNN-Attention model achieves flawless binary classification, attaining 100\% accuracy and a 0\% false positive rate, indicating no misclassifications. For multiclass classification, as shown in Figure \ref{cm}, the model achieves high accuracy across attack types, with 100\% for normal traffic, 98\% for DDoS, and 96\% for Malware. The model’s high diagonal accuracy suggests strong differentiation between attack types, while low off-diagonal misclassifications, such as 4\% Malware instances labeled as DDoS, reinforce its suitability for network security.



\begin{figure}
\centering
\includegraphics[scale=0.25]{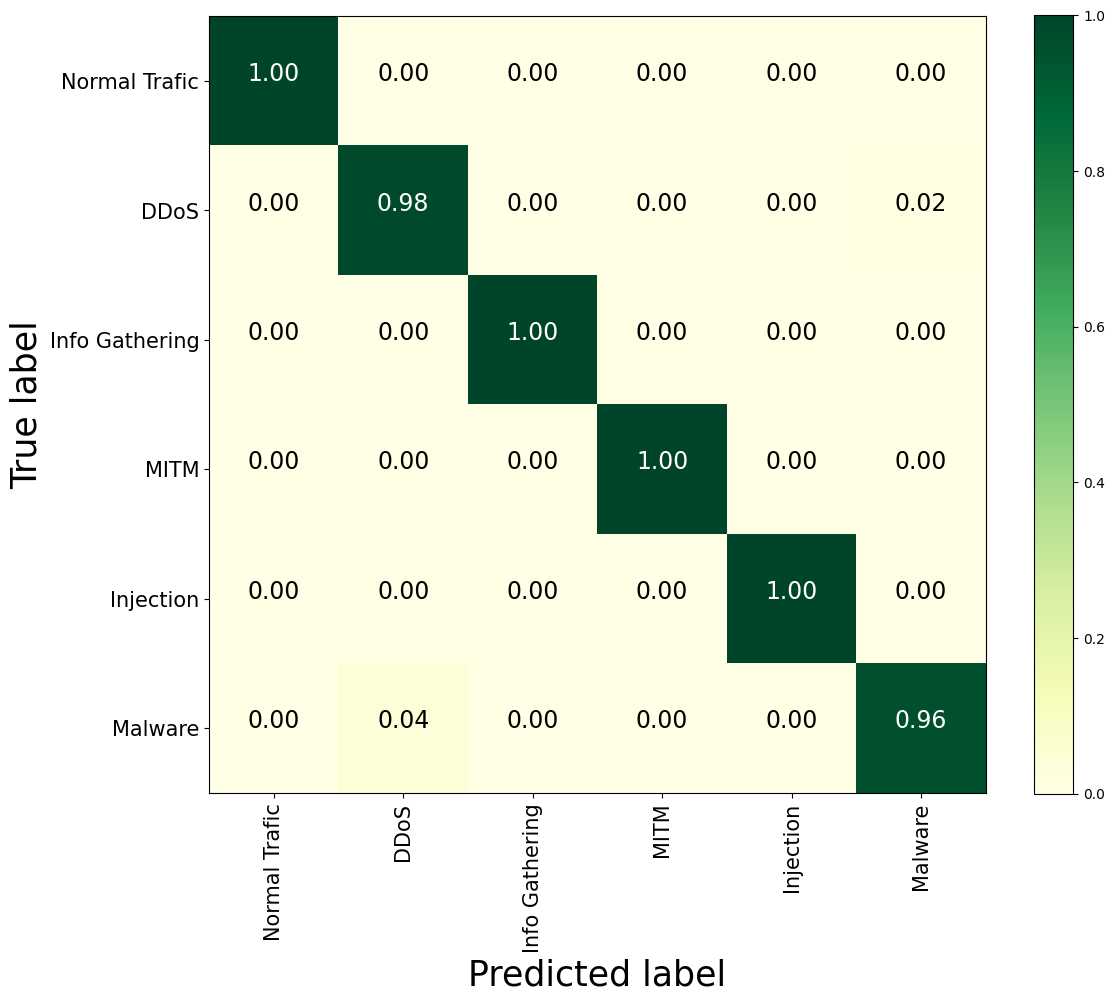}
\caption{Confusion matrix of the LSTM-CNN-Attention model for multiclass classification.}
\label{cm}
\end{figure}

\noindent\textbf{\textit{{- Receiver operating characteristics (ROC):}}} The ROC curve for binary classifications demonstrates perfect classification for both classes (attacks and normal traffic). 
However, the ROC curve in Figure \ref{fig8} shows that the model is performing very well in classifying all six classes. The area under the curve (AUC) for all classes is 100\%, which indicates that the model is able to perfectly distinguish between the classes. This is a very good result and suggests that the model is well-suited for the task of multi-class attack classification.

\begin{figure}
\centering
\includegraphics[scale=0.5]{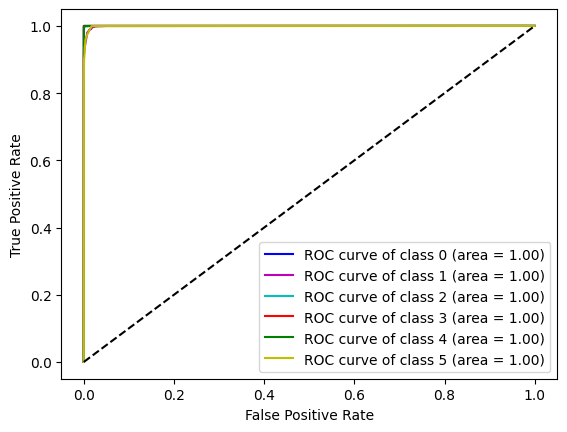}
\caption{ROC curve of the LSTM-CNN-Attention model for multiclass classification.}
\label{fig8}
\end{figure}

\noindent \textbf{\textit{{- Comparison with state-of-the-art:}}}
Table \ref{tab4} compares the performance of the proposed LSTM-CNN-Attention model with other state-of-the-art models for binary and multiclass classification considering various metrics. The table shows that the proposed LSTM-CNN-Attention model achieves the best performance in terms of accuracy, precision, recall, and F1-score for both binary and multiclass classification. The model also has a very low loss. The proposed LSTM-CNN-Attention model also shows the lowest FPR compared to the other methods, with an FPR of 0.002\% for multiclass classification task, demonstrating its superior performance in minimizing FP. This indicates that the model is highly accurate in identifying TP and minimizing the risk of misclassifying negative instances.

\begin{table}[h]
\caption{Comparison to state-of-the-art methods for binary and mult-iclass classification conducted on Edge-IIoTset dataset.}
\label{tab4}
\scriptsize
\begin{tabular}{m{0.3cm}m{1.5cm}m{0.4cm}m{0.5cm}m{0.5cm}m{0.4cm}m{0.4cm}m{0.3cm}m{0.4cm}}
\hline
Work  & Model  & Classif.& Acc. (\%) & Loss & Pr (\%) & Rc (\%) & F1 (\%) & FPR (\%)\\
\hline 
\cite{tareq2022analysis}  & Inception time  & Binary \newline Multi & \ding{55} \newline 94.94 & \ding{55} & \ding{55} & \ding{55} & \ding{55} & \ding{55}\\[1mm]

\cite{ferrag2022edge}  & DNN  & Binary \newline Multi & 99.99 \newline 96.01 & \ding{55} & \ding{55} & \ding{55} & \ding{55} & \ding{55}\\[1mm]

\cite{saadouni2023secure}  & CNN-GRU & Binary \newline Multi & 100 \newline 98.68 & \ding{55} \newline \ding{55}  & 100 \newline 91.89 & \ding{55} \newline \ding{55} & \ding{55} \newline \ding{55} & 0 \newline 0.7\\[0.6mm]

\cite{javeed2023intrusion} & BiGRU-LSTM & Binary \newline Multi& \ding{55} \newline 98.32 & \ding{55} \newline \ding{55}  & \ding{55} \newline 98.78 & \ding{55} \newline 97.22 & \ding{55} \newline \ding{55} & \ding{55} \newline \ding{55} \\[0.6mm]

\textbf{Our}  & \textbf{LSTM-CNN-Attention} &  \textbf{Binary \newline Multi} & \textbf{100 \newline 99.04} & \textbf{$\approx 0$ \newline 0.0220} & \textbf{100 \newline 99.05} & \textbf{100 \newline 99.04} & \textbf{100 \newline 99.04} & \textbf{0 \newline 0.002}\\

\hline
\end{tabular}

\end{table}

\noindent \textbf{\textit{{- Computational Complexity:}}} For real-time \ac{IDS}, testing (or inference) time is more critical than training time. In our proposed method,the  training takes 1509 seconds over 19 epochs (approximately 79 seconds per epoch). However, the testing time is extremely fast at 0.0001 seconds per instance, making it highly suitable for real-time detection. This rapid inference capability allows the model to analyze and classify incoming data almost instantaneously, ideal for real-time \ac{IDS} applications.



\section{Conclusion}
\label{sec5}
In this study, we developed and evaluated an advanced \ac{IDS} for IIoT environments using a hybrid LSTM-CNN-Attention architecture. The model effectively detected and classified cyberattacks in both binary and multiclass tasks using the Edge-IIoTset dataset. We systematically compared various DL models to identify the most effective approach for intrusion detection in complex and dynamic IIoT networks. The experimental results demonstrated that our proposed LSTM-CNN-Attention model significantly outperformed other tested models across all evaluated metrics, achieving near-perfect detection rates. 

Future research could overcome some limitations such as explore performance on more complex datasets, incorporate transfer learning to improve adaptability to new scarce attack patterns and explore optimizations such as lightweight architectures, model compression, federated learning for privacy preserving \cite{himeur2023federated},  and hardware acceleration to mitigate these impacts while retaining detection accuracy.


\section*{Acknowledgment}

The authors are supported by DGRSTD (PRFU-A25N01UN260120230001) 

\begin{scriptsize}
\balance
\bibliographystyle{IEEEtran}
\bibliography{references.bib}    

\begin{thebibliography}{10}
\providecommand{\url}[1]{#1}
\csname url@samestyle\endcsname
\providecommand{\newblock}{\relax}
\providecommand{\bibinfo}[2]{#2}
\providecommand{\BIBentrySTDinterwordspacing}{\spaceskip=0pt\relax}
\providecommand{\BIBentryALTinterwordstretchfactor}{4}
\providecommand{\BIBentryALTinterwordspacing}{\spaceskip=\fontdimen2\font plus
\BIBentryALTinterwordstretchfactor\fontdimen3\font minus \fontdimen4\font\relax}
\providecommand{\BIBforeignlanguage}[2]{{%
\expandafter\ifx\csname l@#1\endcsname\relax
\typeout{** WARNING: IEEEtran.bst: No hyphenation pattern has been}%
\typeout{** loaded for the language `#1'. Using the pattern for}%
\typeout{** the default language instead.}%
\else
\language=\csname l@#1\endcsname
\fi
#2}}
\providecommand{\BIBdecl}{\relax}
\BIBdecl

\bibitem{mishra2021internet}
N.~Mishra and S.~Pandya, ``Internet of things applications, security challenges, attacks, intrusion detection, and future visions: A systematic review,'' \emph{IEEE Access}, vol.~9, pp. 59\,353--59\,377, 2021.

\bibitem{aldhaheri2023deep}
A.~Aldhaheri, F.~Alwahedi, M.~A. Ferrag, and A.~Battah, ``Deep learning for cyber threat detection in {IoT} networks: A review,'' \emph{Internet of Things and Cyber-Physical Systems}, 2023.

\bibitem{kumari2023comprehensive}
P.~Kumari and A.~K. Jain, ``A comprehensive study of {DDoS} attacks over {IoT} network and their countermeasures,'' \emph{Computers \& Security}, vol. 127, p. 103096, 2023.

\bibitem{aguru2024lightweight}
A.~D. Aguru and S.~B. Erukala, ``A lightweight multi-vector {DDoS} detection framework for {IoT}-enabled mobile health informatics systems using deep learning,'' \emph{Information Sciences}, vol. 662, p. 120209, 2024.

\bibitem{mondal2020secure}
B.~Mondal and T.~Mandal, ``A secure image encryption scheme based on genetic operations and a new hybrid pseudo random number generator,'' \emph{Multimedia Tools and Applications}, vol.~79, no.~25, pp. 17\,497--17\,520, 2020.

\bibitem{gueriani2024enhancing}
A.~Gueriani, H.~Kheddar, and A.~C. Mazari, ``Enhancing {IoT} security with cnn and lstm-based intrusion detection systems,'' in \emph{2024 6th International Conference on Pattern Analysis and Intelligent Systems (PAIS)}.\hskip 1em plus 0.5em minus 0.4em\relax IEEE, 2024, pp. 1--7.

\bibitem{jaidka2020evolution}
H.~Jaidka, N.~Sharma, and R.~Singh, ``Evolution of {IoT} to {IIoT}: Applications \& challenges,'' in \emph{Proc. Int. Conf. Innovative Comput. Commun. (ICICC)}, 2020.

\bibitem{sithungu2022gaainet}
S.~P. Sithungu and E.~M. Ehlers, ``Gaainet: A generative adversarial artificial immune network model for intrusion detection in industrial {IoT} systems,'' \emph{J. Adv. Inf. Technol.}, vol.~13, no.~5, 2022.

\bibitem{wang2023securing}
M.~Wang, N.~Yang, and N.~Weng, ``Securing a smart home with a transformer-based iot intrusion detection system,'' \emph{Electronics}, vol.~12, no.~9, p. 2100, 2023.

\bibitem{10729241}
H.~Kheddar, D.~W. Dawoud, A.~I. Awad, Y.~Himeur, and M.~K. Khan, ``Reinforcement-learning-based intrusion detection in communication networks: A review,'' \emph{IEEE Communications Surveys \& Tutorials}, pp. 1--46, 2024.

\bibitem{do2023using}
N.~V. Dalarmelina, P.~Arora, B.~Kaur, R.~I. Meneguette, and M.~A. Teixeira, ``Using ml and dl algorithms for intrusion detection in the industrial internet of things,'' in \emph{AI, Machine Learning and Deep Learning}.\hskip 1em plus 0.5em minus 0.4em\relax CRC Press, 2023, pp. 243--256.

\bibitem{altunay2023hybrid}
H.~C. Altunay and Z.~Albayrak, ``A hybrid cnn+ lstm-based intrusion detection system for industrial {IoT} networks,'' \emph{Engineering Science and Technology, an International Journal}, vol.~38, p. 101322, 2023.

\bibitem{ullah2023abdnn}
S.~Ullah, W.~Boulila, A.~Koubaa, Z.~Khan, and J.~Ahmad, ``{ABDNN-IDS}: Attention-based deep neural networks for intrusion detection in industrial {IoT},'' in \emph{2023 IEEE 98th Veh. Technol. Conf. (VTC2023-Fall)}.\hskip 1em plus 0.5em minus 0.4em\relax IEEE, 2023, pp. 1--5.

\bibitem{kheddar2024transformers}
H.~Kheddar, ``Transformers and large language models for efficient intrusion detection systems: A comprehensive survey,'' \emph{arXiv preprint arXiv:2408.07583}, 2024.

\bibitem{habchi2024machine}
Y.~Habchi, H.~Kheddar, Y.~Himeur, A.~Boukabou, A.~Chouchane, A.~Ouamane, S.~Atalla, and W.~Mansoor, ``Machine learning and vision transformers for thyroid carcinoma diagnosis: A review,'' \emph{arXiv preprint arXiv:2403.13843}, 2024.

\bibitem{djeffal2023automatic}
N.~Djeffal, H.~Kheddar, D.~Addou, A.~C. Mazari, and Y.~Himeur, ``Automatic speech recognition with bert and ctc transformers: A review,'' in \emph{2023 2nd Int. Conf. on Electronics, Energy and Measurement (IC2EM)}, vol.~1.\hskip 1em plus 0.5em minus 0.4em\relax IEEE, 2023, pp. 1--8.

\bibitem{alshehri2024self}
M.~S. Alshehri, O.~Saidani, F.~S. Alrayes, S.~F. Abbasi, and J.~Ahmad, ``A self-attention-based deep convolutional neural networks for {IIoT} networks intrusion detection,'' \emph{IEEE Access}, 2024.

\bibitem{de2022hybrid}
E.~M. de~Elias, V.~S. Carriel, G.~W. De~Oliveira, A.~L. Dos~Santos, M.~Nogueira, R.~H. Junior, and D.~M. Batista, ``A hybrid cnn-lstm model for iiot edge privacy-aware intrusion detection,'' in \emph{2022 IEEE Latin-American Conference on Communications (LATINCOM)}.\hskip 1em plus 0.5em minus 0.4em\relax IEEE, 2022, pp. 1--6.

\bibitem{sharma2024explainable}
B.~Sharma, L.~Sharma, C.~Lal, and S.~Roy, ``Explainable artificial intelligence for intrusion detection in {IoT} networks: A deep learning based approach,'' \emph{Expert Systems with Applications}, vol. 238, p. 121751, 2024.

\bibitem{ullah2023magru}
S.~Ullah, W.~Boulila, A.~Koubaa, and J.~Ahmad, ``Magru-ids: A multi-head attention-based gated recurrent unit for intrusion detection in {IIoT} networks,'' \emph{IEEE Access}, 2023.

\bibitem{vaswani2017attention}
A.~Vaswani, ``Attention is all you need,'' \emph{Advances in Neural Information Processing Systems}, 2017.

\bibitem{kheddar2024automatic}
H.~Kheddar, M.~Hemis, and Y.~Himeur, ``Automatic speech recognition using advanced deep learning approaches: A survey,'' \emph{Information Fusion}, p. 102422, 2024.

\bibitem{ferrag2022edge}
M.~A. Ferrag, O.~Friha, D.~Hamouda, L.~Maglaras, and H.~Janicke, ``Edge-{IIoT}set: A new comprehensive realistic cyber security dataset of {IoT} and {IIoT} applications for centralized and federated learning,'' \emph{IEEE Access}, vol.~10, pp. 40\,281--40\,306, 2022.

\bibitem{gueriani2023deep}
A.~Gueriani, H.~Kheddar, and A.~C. Mazari, ``Deep reinforcement learning for intrusion detection in {IoT}: A survey,'' in \emph{2023 2nd Int. Conf. on Electronics, Energy and Measurement (IC2EM)}, vol.~1.\hskip 1em plus 0.5em minus 0.4em\relax IEEE, 2023, pp. 1--7.

\bibitem{kheddar2024deep}
H.~Kheddar, M.~Hemis, Y.~Himeur, D.~Meg{\'\i}as, and A.~Amira, ``Deep learning for steganalysis of diverse data types: A review of methods, taxonomy, challenges and future directions,'' \emph{Neurocomputing}, p. 127528, 2024.

\bibitem{dunn2020robustness}
C.~Dunn, N.~Moustafa, and B.~Turnbull, ``Robustness evaluations of sustainable machine learning models against data poisoning attacks in the internet of things,'' \emph{Sustainability}, vol.~12, no.~16, p. 6434, 2020.

\bibitem{kheddar2023deep}
H.~Kheddar, Y.~Himeur, and A.~I. Awad, ``Deep transfer learning for intrusion detection in industrial control networks: A comprehensive review,'' \emph{J. Netw. Comput. Appl.}, vol. 220, p. 103760, 2023.

\bibitem{tareq2022analysis}
I.~Tareq, B.~M. Elbagoury, S.~El-Regaily, and E.-S.~M. El-Horbaty, ``Analysis of ton-iot, unw-nb15, and edge-iiot datasets using dl in cybersecurity for iot,'' \emph{Applied Sciences}, vol.~12, no.~19, p. 9572, 2022.

\bibitem{saadouni2023secure}
R.~Saadouni, A.~Khacha, Y.~Harbi, C.~Gherbi, S.~Harous, and Z.~Aliouat, ``Secure {IIoT} networks with hybrid cnn-gru model using edge-{IIoT}set,'' in \emph{2023 15th Int. Conf. on Innovations in Inf. Technol. (IIT)}.\hskip 1em plus 0.5em minus 0.4em\relax IEEE, 2023, pp. 150--155.

\bibitem{javeed2023intrusion}
D.~Javeed, T.~Gao, M.~S. Saeed, and P.~Kumar, ``An intrusion detection system for edge-envisioned smart agriculture in extreme environment,'' \emph{IEEE Internet of Things Journal}, 2023.

\bibitem{himeur2023federated}
Y.~Himeur, I.~Varlamis, H.~Kheddar, A.~Amira, S.~Atalla, Y.~Singh, F.~Bensaali, and W.~Mansoor, ``Federated learning for computer vision,'' \emph{arXiv preprint arXiv:2308.13558}, 2023.

\end{thebibliography}
\end{scriptsize}

\end{document}